\documentclass[aps,prc,preprint,showpacs,superscriptaddress,floatfix,raggedbottom]{revtex4-1}
\usepackage{amsmath, amsthm, amssymb}
\usepackage{graphics}
\usepackage{graphicx}
\usepackage{xcolor}
\usepackage{epsfig}
\usepackage{gensymb}        
\usepackage{float}
\usepackage{multirow}   
\usepackage{subfig}    
\usepackage{array}
\begin{document}
\title{Elastic scattering measurements for the \textsuperscript{10}C + \textsuperscript{208}Pb system at E$_{\textnormal{lab}}$~=~66~MeV}

\author{R.~Linares}
\email[corresponding author: ]{rlinares@id.uff.br}
\affiliation{Instituto de F\'isica, Universidade Federal Fluminense, 24210-340, Niter\'oi, Rio de Janeiro, Brazil}

\author{Mandira~Sinha}
\affiliation{Department of Physics, Bose Institute, 93/1 Acharya Prafulla Chandra Road, Kolkata-700009, W.B, India}

\author{E.~N.~Cardozo}
\affiliation{Instituto de F\'isica, Universidade Federal Fluminense, 24210-340, Niter\'oi, Rio de Janeiro, Brazil}
\affiliation{Instituto de F\'isica, Universidade de S\~ao Paulo, 05508-090 S\~ao Paulo, Brazil}

\author{V.~Guimar\~aes}
\affiliation{Instituto de F\'isica, Universidade de S\~ao Paulo, 05508-090 S\~ao Paulo, Brazil}

\author{G.~V.~Rogachev}
\affiliation{Department of Physics and Astronomy, Texas A \& M University, 77843 TX}
\affiliation{Cyclotron Institute, Texas A \& M University, 77843 TX}
\affiliation{Nuclear Solutions Institute, Texas A \& M University, 77843 TX}

\author{J.~Hooker}
\affiliation{Department of Physics and Astronomy, Texas A \& M University, 77843 TX}
\affiliation{Cyclotron Institute, Texas A \& M University, 77843 TX}

\author{E.~Koshchiy}
\author{T.~Ahn}
\affiliation{Cyclotron Institute, Texas A \& M University, 77843 TX}

\author{C.~Hunt}
\affiliation{Department of Physics and Astronomy, Texas A \& M University, 77843 TX}
\affiliation{Cyclotron Institute, Texas A \& M University, 77843 TX}

\author{H.~Jayatissa}
\altaffiliation[Present address: ]{ Physics Division, Argonne National Laboratory, Argonne, IL 60439, USA.}
\affiliation{Department of Physics and Astronomy, Texas A \& M University, 77843 TX}
\affiliation{Cyclotron Institute, Texas A \& M University, 77843 TX}

\author{S.~Upadhyayula}
\affiliation{Department of Physics and Astronomy, Texas A \& M University, 77843 TX}
\affiliation{Cyclotron Institute, Texas A \& M University, 77843 TX}

\author{B.~Roeder}
\author{A.~Saastomoinen}
\affiliation{Cyclotron Institute, Texas A \& M University, 77843 TX}

\author{J.~Lubian}
\affiliation{Instituto de F\'isica, Universidade Federal Fluminense, 24210-340, Niter\'oi, Rio de Janeiro, Brazil}

\author{M.~Rodr\'iguez-Gallardo}
\affiliation{Departamento de F\'isica At\'omica, Molecular y Nuclear, Facultad de F\'isica, Universidad de Sevilla, Apartado 1065, E-41080 Sevilla, Spain\\
and Instituto Carlos I de F\'{i}sica Te\'{o}rica y Computacional, Universidad de Sevilla, Spain}

\author{J.~Casal}
\affiliation{Dipartimento di Fisica e Astronomia ``G. Galilei'', Universit\`a degli studi di Padova, and INFN - Sezione di Padova, Via Marzolo 8, I-35131 Padova, Italy}

\author{K. C. C.~Pires}
\affiliation{Instituto de F\'isica, Universidade de S\~ao Paulo, 05508-090  S\~ao Paulo, Brazil}

\author{M.~Assun\c c\~ao}
\affiliation{Departamento F\'isica, Universidade Federal de S\~ao Paulo, Campus Diadema, 09913-030, Diadema, S\~ao Paulo, Brazil}

\author{Y. Penionzhkevich}
\author{S. Lukyanov}
\affiliation{Joint Institute for Nuclear Research, Dubna. Moscow region, 141980}

\date{\today}

\begin{abstract} 
\textbf{Background:}  The influence of halo structure of $^6$He, $^8$B, $^{11}$Be and $^{11}$Li nuclei in several mechanisms such as direct reactions and fusion is already established, although not completely understood. The influence of the $^{10}$C Brunnian structure is less known.

\textbf{Purpose:} To investigate the influence of the cluster configuration of $^{10}$C on the elastic scattering at an energy close to the Coulomb barrier.

\textbf{Methods:} We present experimental data for the elastic scattering of the $^{10}$C+$^{208}$Pb system at $E_{\rm lab}$ = 66~MeV. The data are compared to the three- and the four-body continuum-discretized coupled-channels calculations assuming $^9$B+$p$, $^6$Be+$\alpha$ and $^8$Be+$p$+$p$ configurations.

\textbf{Results:} The experimental angular distribution of the cross sections shows the suppression of the Fresnel peak that is reasonably well reproduced by the  continuum-discretized coupled-channels calculations. However, the calculations underestimate the cross sections at backward angles. Couplings to continuum states represent a small effect.  

\textbf{Conclusions:} The cluster configurations of $^{10}$C assumed in the present work are able to describe some of the features of the data. In order to explain the data at backward angles, experimental data for the breakup and an extension of theoretical formalism towards a four-body cluster seem to be in need to reproduce the measured angular distribution.

\end{abstract}

\pacs{}

\maketitle

\section{Introduction}
\label{Intro}

The investigation of nuclear structure and reaction mechanisms induced by radioactive nuclei is a subject of considerable interest. Detailed reviews on this subject are presented in refs.~\cite{KGA16, CGD15, CGL20}. Some of the weakly-bound nuclei are called exotic as they exhibit low binding energies and halo structure, in which the valence nucleon(s) orbits a compact inert core and therefore forms an extended matter distribution \cite{THH85, TSK13}. These features of the exotic nuclei manifest themselves in nuclear collisions mainly at energies around the Coulomb barrier. At these energies, a strong interplay between elastic scattering and other processes, such as breakup and transfer reactions, is expected. For instance, the low binding energy produces a decoupling between the valence nucleon(s) and the core during the collision which might cause suppression of the Fresnel peak (Coulomb rainbow) in the angular distributions \cite{KGA16}.

Also, the decoupling to the continuum of the single-particle or cluster motion of the valence nucleon(s) with respect to the core has motivated the application of reaction models such as continuum-discretized coupled-channels (CDCC) \cite{AIK87}, in which the coupling between the elastic and the breakup channel is explicitly taken into account. The CDCC calculations have been applied to study the elastic scattering of exotic neutron-rich nuclei, such as $^6$He, $^{11}$Li and $^{11}$Be, on several medium to heavy target nuclei and at energies around the Coulomb barrier, \cite{MPR14, PFA04, FLP10, SEA08,ASG11, AVQ11, PRS10, PSM12, AAA09, PBM17, MSR06}. In some of these works, the $^{11}$Be has been described as $^{10}$Be core and a valence neutron while $^6$He has been described as a $^4$He core and a paired di-neutron system. This is referred as three-body CDCC (3b-CDCC) framework. An extension of this framework allows to treat the projectile as a core plust two valence particles configuration, simply referred as the four-body CDCC (4b-CDCC) calculations \cite{RAG09}. The 4b-CDCC formalism has been applied to study reactions with $^6$He \cite{MPR14}, $^9$Be \cite{CRA15, ACR18} and $^{11}$Li \cite{CFR12, FCR13}, in which the $^6$He and $^{11}$Li are treated as a core+$n$+$n$ and $^9$Be as a $\alpha$+$\alpha$+$n$ configurations. These cluster configurations are quite important to describe the elastic scattering data.

Moving towards the proton-rich side, one of the most studied nucleus is $^8$B. This nucleus has a very weakly bound valence proton with separation energy $S_p=0.138$ MeV. The centrifugal barrier between the valence proton, in the $p$-orbital, and the core inhibits the wider radial extension. Therefore the halo structure is relatively less pronounced when compared to the neutron-halo nuclei. Elastic scattering for the $^8$B+$^{58}$Ni system was investigated in Refs.~\cite{AML09, LCA09, KMB10}. The observed large reaction cross sections has been interpreted as an indication of halo configuration for this nucleus. This is  also important to describe the $^{8}$B $\rightarrow$ $^{7}$Be + $p$ breakup cross sections at sub-Coulomb energy \cite{GKP00}.

The effect of the breakup channel on the elastic scattering can be studied in collisions with heavier targets due to the increasing predominance of the long-range Coulomb interaction over the nuclear interaction, that induces a strong dipole response. In Ref.~\cite{YWW13} it has been reported measurements of the elastic scattering in the $^{8}$B + $^{208}$Pb system at E$_{\textnormal{lab}}$ = 170~MeV, that corresponds to about three times the Coulomb barrier height. The effect of breakup on the elastic scattering is small at this bombarding energy but becomes relevant at energies around the Coulomb barrier, as reported in Refs.~\cite{RLC16, YLP16}. Recently, Mazzocco {\it et al.,} reported a measurement of the elastic scattering for the same $^{8}$B + $^{208}$Pb system  at E$_{\textnormal{lab}}$ = 50~MeV~\cite{MKB19}. The data indicate a strong suppression of the Fresnel peak and a large reaction cross section. However, the 3b-CDCC calculations do not provide a full description of the data.

The $^{10}$C is another interesting proton-rich nucleus due to its Brunnian (super-Borromean) structure where the four interconnected rings are associated to the interactions between $\alpha$+$\alpha$+$p$+$p$ \cite{CAA08}. The removal of one particle breaks the remaining system apart. This nucleus can decay by three possible channels: $^8$Be + $p$ + $p$ , $^{9}$B + $p$  and $^{6}$Be+$\alpha$ with binding energies of 3.821, 4.006 and 5.101 MeV, respectively. Since $^8$Be, $^9$B and $^6$Be are unbound systems, the final state is the $\alpha+\alpha+p+p$ system. Breakup measurement of the $^{10}$C on $^{12}$C at 33 MeV/A indicates that the final state is predominantly fed by the $^9$B+$p$ and $^8$Be+$p$+$p$ channels \cite{CAA08}. 

Measurementes of the elastic scattering has been recently performed for the $^{10}$C+$^{58}$Ni at E$_{\textnormal{lab}}$ = 35.3~MeV \cite{GCS19}. Couplings to the first excited state in $^{10}$C ($2_1^+$) has been quite important in the description of cross sections at backward angles, indicating the importance of the large deformation for this nucleus. Analysis within the 3b-CDCC, assuming the $^{9}$B + $p$  and $^{6}$Be+$\alpha$ configurations, also indicates a strong cluster configuration in the $^{10}$C. Elastic measurements of $^{10}$C on  $^{208}$Pb target were performed at 226 and 256~MeV \cite{YWW14}, well above the Coulomb barrier, and the angular distributions exhibit a typical Fresnel-like shape.

In this work we present new experimental data for the elastic scattering of $^{10}$C on $^{208}$Pb target measured at E$_{\textnormal{lab}} = 66$~MeV, which corresponds to an energy close to the Coulomb barrier (V$_{\textnormal{B}}$ = 61.3~MeV in the laboratory framework). The theoretical analyses are carried out within the optical model, to determine the reaction cross section, and the coupled channel, 3b- and 4b-CDCC formalisms to assess the contribution of individual reaction channels. The paper is organized as follows: the experimental details and the theoretical analysis are discussed in sections~\ref{exp} and~\ref{theor}, respectively. The conclusions are given in section~\ref{conc}.


\section{\label{exp}Experimental details}

The experiment was performed at the Cyclotron Institute of Texas A\&M University. The radioactive $^{10}$C beam was produced by the charge-exchange  \textit{p}($^{10}$B,$^{10}$C)\textit{n} reaction. The primary 9.6 MeV/u $^{10}$B beam was delivered by the K500 superconducting cyclotron and impinged on a H$_2$ gas cell cooled to 77 K. The produced $^{10}$C ions were selected by the Momentum Achromatic Recoil Spectrometer (MARS) and directed to the scattering chamber \cite{TAG02}. The production rate of the radioactive beam was monitored as the ratio between the thin plastic scintillators counting, at the entrance of the scattering chamber, and the primary beam current collect at the gas cell. The average production rate was $5\times10^{3}$~pps for a 300~nA incident $^{10}$B$^{+5}$ beam. The obtained purity for $^{10}$C beam was better than 95\%.

\begin{figure}[tb!]
\centering
\graphicspath{{figuras/}}
\includegraphics[width=0.45\textwidth]{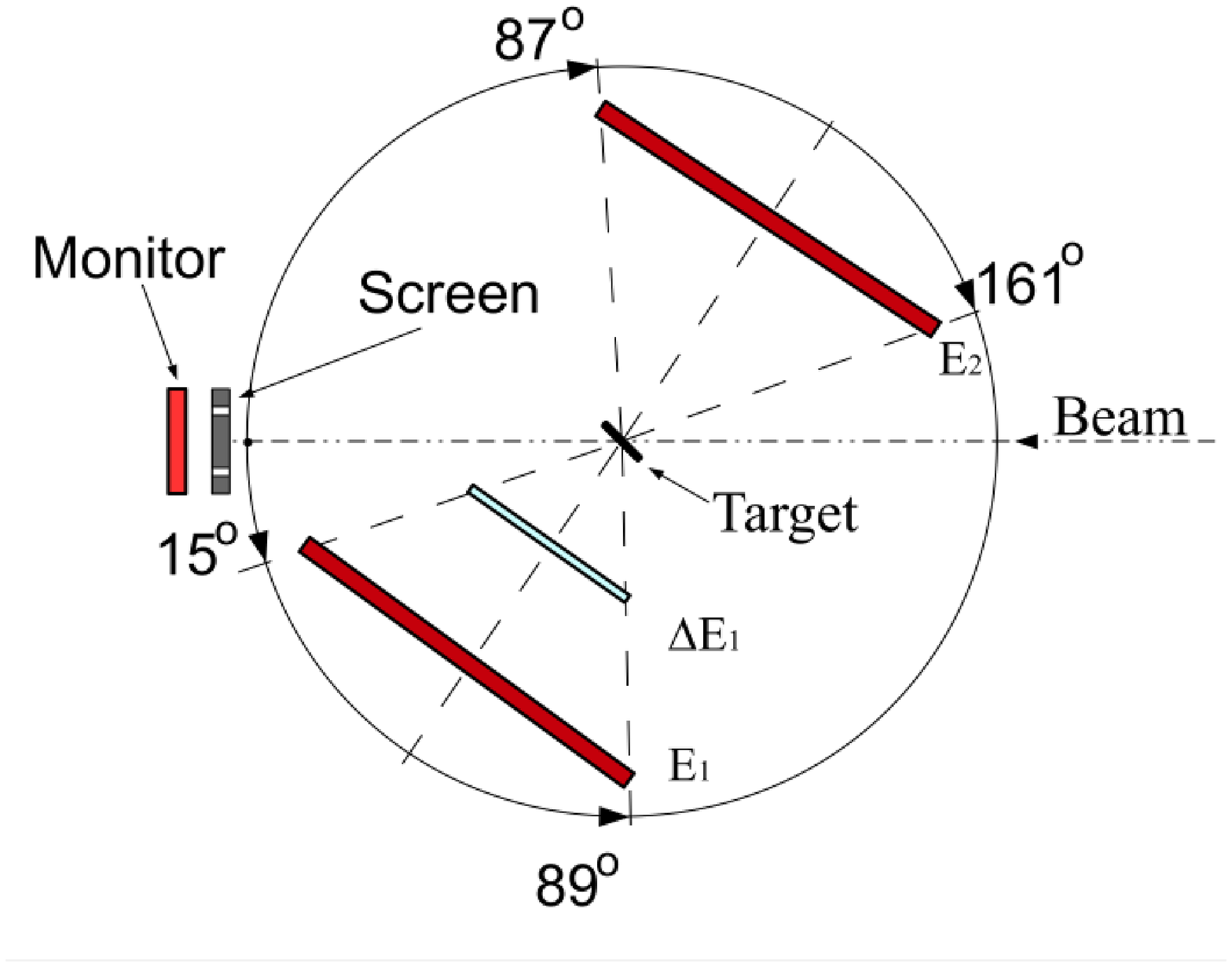}
\caption{(Color online) Sketch of the detection system for the measurements, composed by one $\Delta E-E$ telescope, placed at forward angles, and a E-detector at backward angles. The E-detectors are $128 \times 128$ DSSSD and the $\Delta E$ is a $16 \times 16$ thin DSSSD. A third DSSSD (monitor) was placed at $0\degree$ to monitor the beam axis.} 
\label{fig:setup}
\end{figure}

The schematic diagram of the experimental arrangement in the scattering chamber is shown in Fig.~\ref{fig:setup}. The target consisted of two isotopically enriched $^{208}$Pb foils, 2.37~mg/cm$^2$ thick, evaporated on a 0.08~mg/cm$^2$ thick carbon backing. These targets were mounted together on the target frame. The detection system consisted of a double-sided silicon strip detectors (DSSSD) $\Delta$E-E$_1$ telescope placed at forward scattering angles and a single DSSSD (E$_2$) at backward scattering angles. The E$_1$ and E$_2$ detectors have an active area of 97.2~mm $\times$ 97.2~mm, 500~$\mu$m thick and with front and back sides segmented in 128 strips. Both detectors were placed 100~mm away from the target position. The $\Delta$E detector of the telescope has an active area of 49.5~mm $\times$ 49.5~mm,  38~$\mu$m thick, and had front and back sides segmented in 16 strips. This detector was placed 50~mm away from the target position. These detectors were mounted in a diamond-like geometry to allow for a wide angular coverage. In this configuration, the telescope covered the scattering angles between $\theta_{\rm lab} = 15\degree$ and $89\degree$ whereas the E$_2$ covered the angular range from $\theta_{\rm lab} = 87\degree$ to $161\degree$.

An extra $16 \times 16$ DSSSD detector was placed at  $0\degree$, downstream the beam and at the back of the chamber to monitor the radioactive $^{10}$C beam. To avoid damages in this monitor due to direct exposure to the secondary beam, a screen with four 2-millimeter holes, positioned at the vertices of a 20~mm square, was mounted in front of the detector. The relative intensities registered at these positions were used to determine the central position of the secondary beam as well as the $^{10}$C beam spot size, estimated to be 6~mm.

\begin{figure}[tb!]
\centering
\graphicspath{{figuras/}}
\includegraphics[width=0.5\textwidth]{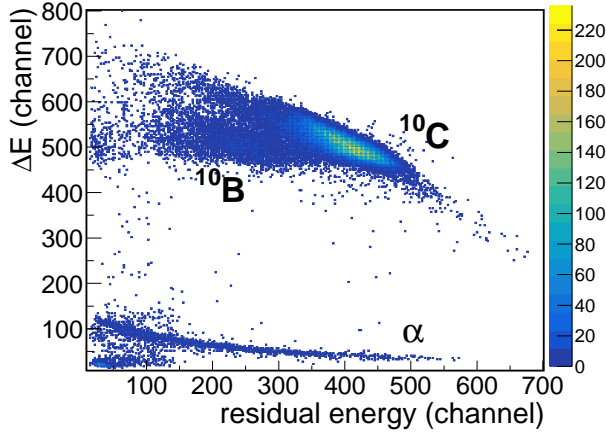}
\caption{(Color online) The $\Delta E-E$ plot for particle identification. A small fraction of \textsuperscript{10}B is observed due to sparks in the Wien filter of MARS separator during the acquisition runs.} 
\label{fig:identification_spectra}
\end{figure}

The two-dimensional particle identification spectrum, $\mathit{\Delta E-E\textsubscript{\rm res.}}$, obtained with the telescope is shown in Fig.~\ref{fig:identification_spectra}. The elastically scattered $^{10}$C ions are well separated from the $^{10}$B ions. The small amount of $^{10}$B ions went through the MARS beam line due to sparks in the Wien filter during the acquisition runs. The observed light particles ($\alpha$ and protons) are from fusion-evaporation and break-up reactions. However, according to calculations of the fusion-evaporation process performed with the the \textsc{pace} code \cite{TaB08,Gav80}, there is no significant contribution of \textit{p} and $\alpha$ particles in the energy region of interest of the elastic scattering at backward angles (detector E$_2$).

\begin{figure}[tb!]
\centering
\graphicspath{{figuras/}}
\includegraphics[width=0.95\textwidth]{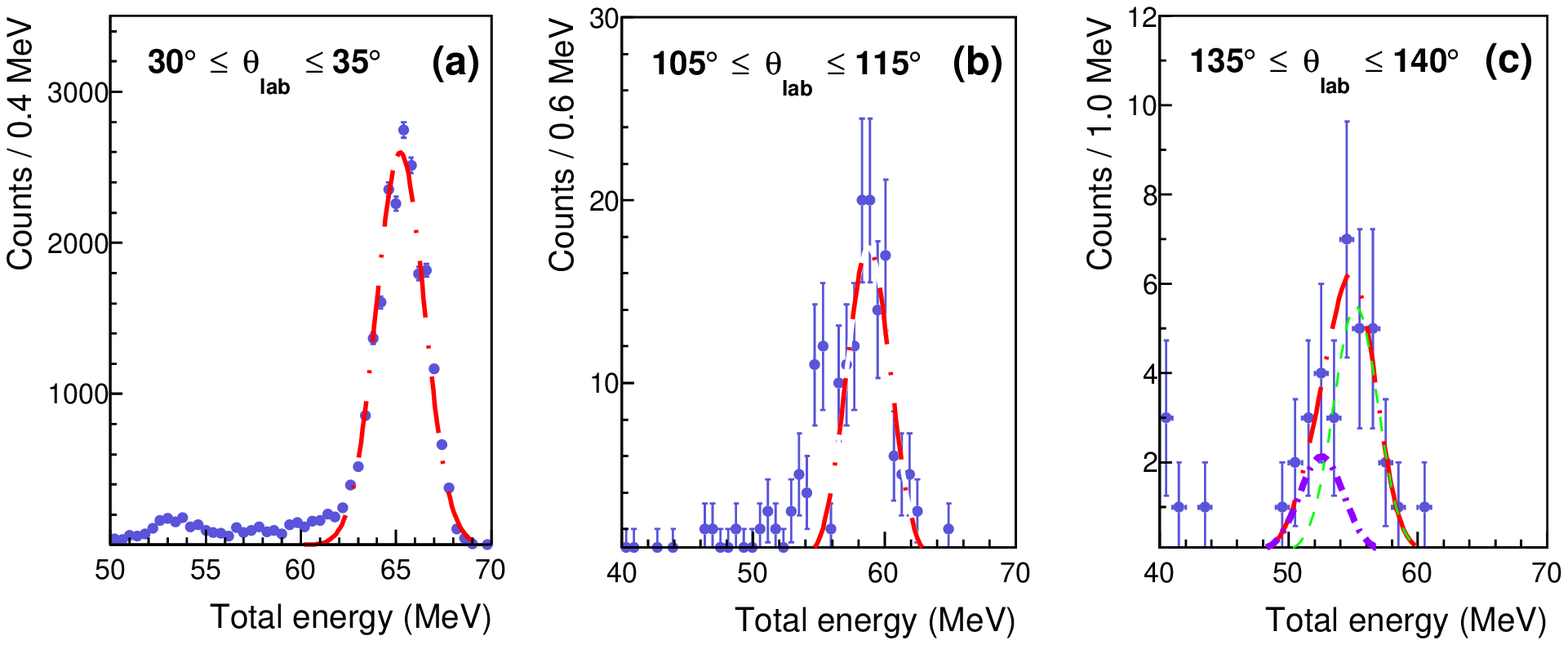}
\caption{(Color online) Example of total energy spectrum for \textsuperscript{10}C particles scattered by \textsuperscript{208}Pb at (a) $30\degree \leq \theta _{\textnormal{lab}} \leq 35\degree$, (b) $105\degree \leq \theta _{\textnormal{lab}} \leq 115\degree$ and (c) $135\degree \leq \theta _{\textnormal{lab}} \leq 140\degree$. In panels (a) and (b), the elastic peak is represented by a single gaussian curve. In panel (c), data points were adjusted to a two-gaussian curve with the same standard deviation. The green dotted and violet dashed-dot curves correspond to elastic and the contribution from $2^+$ state in $^{10}$C and $3^-$ in $^{208}$Pb, respectively.} 
\label{fig:energy-spectra}
\end{figure}

The total energy spectrum (E$_{\textnormal{total}}$ = $\Delta E + E$) at 3 different scattering angles are shown in Fig.~\ref{fig:energy-spectra}. The achieved energy resolution is about 2.9~MeV (Full Width at Half Maximum). This energy resolution does not allow a complete separation of the elastic peak from the inelastic scattering leading to the $3^-_1$ state in $^{208}$Pb (at 2.61~MeV). At the very backward angles ($\theta_{\textnormal{lab.}} > 120\degree$), yields of the elastic contribution were determined considering 2-Gaussian curve fits, being the second Gaussian associate to the inelastic scattering contributions of the $2^+$ state in $^{10}$C and $3^-$ in $^{208}$Pb Therefore, in principle, this is a quasi-elastic scattering measurement. Also, according to calculations, the contribution of inelastic scattering is found to be quite small, as it will be shown in the next sections. Also, the spectra at backward angles show almost no background in the elastic peak region, which corroborate with the absence of evaporated $\alpha$ in the region of interest. Considering the energy losses in the scintillators ($14 \mu$m thick), upstream the scattering chamber, and in the Pb+C target, the averaged energy is 65.7~MeV. In particular, the estimated energy loss in the target is 2.1~MeV.

The yields of \textsuperscript{10}C scattered particles were integrated in angular ranges of $5\degree$ or $10\degree$, depending on the counting statistics. The scattering angle of each detected particle was determined from the coordinates of the strips triggered in the E-detector. The angular resolution is estimated to be about $3\degree$ (in the laboratory framework), where the beam spot size at the target, the relative distances between target holder and E-detectors and the strip width have been taken into account. The solid angles were determined using an alpha source placed at the target position and assuming an uniform emission. The final overall normalization was performed by considering the cross sections for $\theta_{\rm lab} < 45\degree$ as Rutherford cross section.

\begin{figure}[tb!]
\centering
\graphicspath{{figuras/}}
\includegraphics[width=0.5\textwidth]{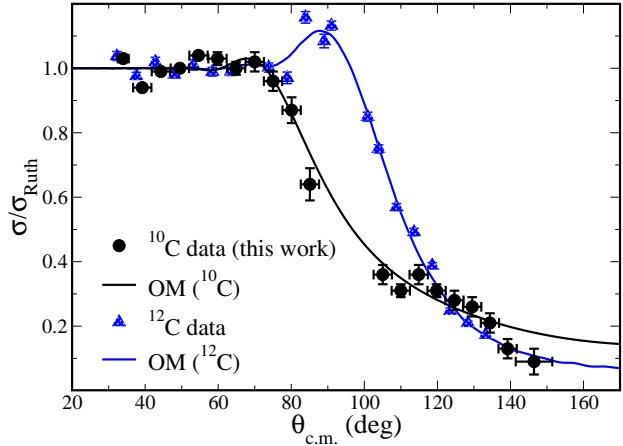}
\caption{(Color online) Elastic scattering angular distribution for the $^{10}$C + $^{208}$Pb system at E$_{\textnormal{lab}}$ = 66~MeV (black circle points). The points in blue triangles correspond to data for $^{12}$C + $^{208}$Pb system at  64.9~MeV from Ref.~\cite{SSK01}. The solid curves were obtained by fitting the parameters of the optical potential.}
\label{fig:AngDistExp}
\end{figure}

The experimental angular distribution of the elastic cross sections is shown in Fig.~\ref{fig:AngDistExp}. The error bars in the experimental cross sections correspond to the statistic uncertainties only. Systematic uncertainties due to solid angles and the beam spot size at the target introduce about 10\% uncertainty to the cross sections. The angular distribution for the elastic scattering of $^{12}$C + $^{208}$Pb system at E$_{\rm lab}$~=~64.9~MeV \cite{SSK01} is also plotted in the figure (blue triangle points). The comparison of both angular distributions highlights the suppression of the Fresnel peak for the $^{10}$C projectile. The solid curves correspond to optical model calculations with fitting parameters and are described in the next section.

\section{\label{theor}Theoretical Analysis}

\subsection{Optical model calculation}

In a first approach, the angular distribution of the elastic scattering is analyzed within the optical model (OM) using a Woods-Saxon (WS) shape for the real and imaginary potentials. The depths, radii and diffuseness parameters were adjusted to minimize the $\chi^2$ parameter between data and the OM cross section. The parameters obtained in the fitting procedure for the $^{12}$C + $^{208}$Pb data at 64.9~MeV \cite{SSK01} were adopted as initial guessing parameters for the $^{10}$C + $^{208}$Pb system. The best values obtained by the fitting procedure ($\chi^2 = 48.2$) are shown in Table~\ref{tab:opticalmodel} along with the corresponding reaction cross section, $\sigma_R = 753$~mb.

\begin{table*} [t]
\caption{Parameters of the optical potentials obtained by fitting of the elastic scattering in the
$^{12}$C + $^{208}$Pb data at E$_{\rm lab}$~=~64.9~MeV and $^{10}$C + $^{208}$Pb data at E$_{\rm lab}$~=~66~MeV. Parameters $V$ and $W$ are in MeV. The $r_V$ and $r_W$ are reduced radii,
using the convention $R_i = r_i \times [A_p^{1/3} + A_t^{1/3}]$~fm and $r_C = 1.3$~fm. The $a_V$ and $a_W$ are the diffuseness, in fm. The reaction cross sections ($\sigma_R$) obtained with the optical potentials are also presented.} 
\centering
\begin{tabular}{c c c c c c c c  }
\hline
projectile & $V$ & $r_V$ & $a_V$ & $W$ & $r_W$ & $a_W$ & $\sigma_{R}$~(mb) \\
\hline
$^{12}$C & 83.7~ & 1.21~ & 0.63~ & 33.7~ & 1.37~ & 0.24~ & 382 \\
$^{10}$C & 82.2~ & 1.19~ & 0.12~ & 17.6~ & 1.60~ & 0.14~ & 753  \\
\hline
\end{tabular}
\label{tab:opticalmodel}
\end{table*}

To compare the $\sigma_R$ for different projectiles on $^{208}$Pb it is necessary to adopt a reduction method which removes the geometrical and charged properties of the systems. There are three different reduction methods for reaction cross section available in the literature, which, following the nomenclature used by Canto \textit{et al.} \cite{CMG15}, are referred as: traditional, simplified traditional \cite{GLP05}, and fusion (reaction) function \cite{CGL09b,SGL09}. Although none of this methods have acceptable performance for a wide variation of systems, they can be reliable for reaction cross sections of similar systems with same target~\cite{CMG15, CGL20}. Here we adopt the fusion (reaction) function method. In this case, the reduced energy ($E_{\rm Red}$) and reduced reaction cross section ($\sigma_{\rm Red}$) are defined as follows:

\begin{equation}
E_{\rm Red} = \frac{E - V_{\rm B}}{\hbar\omega}, \qquad
\sigma_{\rm Red} = \sigma_{\rm R} \left[ \frac{2E}{\hbar\omega
    R_{\rm B}^2} \right].
\end{equation}

In both equations, $R_{\rm B}$, $V_{\rm B}$ and $\hbar\omega$ are the radius, height and width of a parabolic shape adjusted to the Coulomb barrier. Here we considered the double folding S\~ao Paulo potential \cite{CCG02} to obtain these parameters. For the $^{10}$C projectile, the adjusted $R_{\rm B}$, $V_{\rm B}$ and $\hbar\omega$ are 11.3~fm, 58.8~MeV and 5.2~MeV, respectively. Also, it is important to correct the averaged energy for the target thickness, according to Ref.~\cite{FSD15}. Assuming that energies of particles that compose the incident beam has a gaussian distribution, with standard deviation of 0.8~MeV, and the target has an uniform thickness, the effective energy ($E_{\text{eff}}$) is 66.4~MeV. This value has been considered to calculate the corresponding $E_{\rm Red}$.

\begin{figure}[tb!]
\centering
\graphicspath{{figuras/}}
\includegraphics[width=0.5\textwidth]{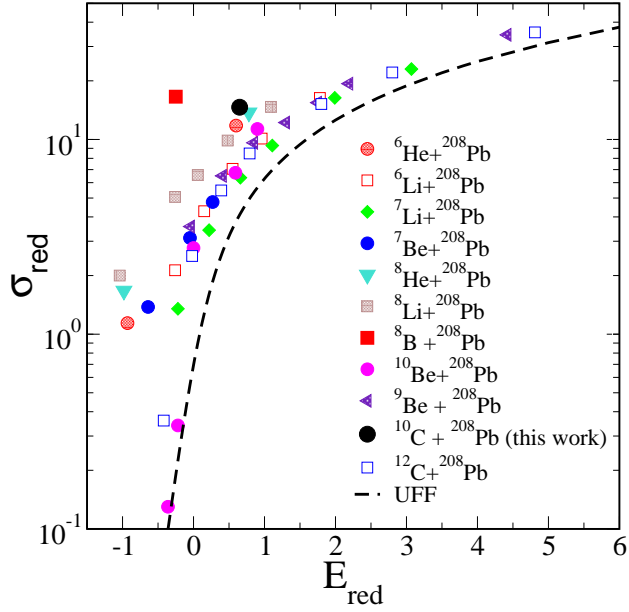}
\caption{(Color online) Reduced reaction cross section for the
several projectiles on $^{208}$Pb target. The universal fusion function (UFF) is also shown in the plot as a lower bound to the reduced reaction cross sections.} 
\label{fig:ReactionReduction}
\end{figure}

In Fig.~\ref{fig:ReactionReduction}, we compared the reduced cross section for the $^{10}$C with other projectiles in $^{208}$Pb target: $^{6,8}$He \cite{ASG11,SEA08, MMS16}, $^{6, 7, 8}$Li \cite{KBC94, KGL02}, $^{9,10}$Be \cite{WFC04, KAB04}, $^{8}$B \cite{MKB19} and $^{12}$C \cite{SSK01}. The dashed curve in the figure corresponds to the Universal Fusion Function (UFF), introduced in Refs.~\cite{CGL09a,CGL09b}. This function is related to the fusion cross sections based on the Wong's equation. It is, thus, a lower bound for the reduced reaction cross sections. The reduced reaction cross section for the $^{10}$C, obtained with the optical model, lies a little above of the weakly-bound $^{6,8}$He and $^{8}$Li projectiles. 

An enhancement of the reduced cross section, compared to the UFF curve, indicates that other reaction channels, driven by large deformation  and/or cluster configuration, give a sizeable contribution. This is actually the case for $^{6}$He and $^8$B, for which transfer and/or breakup are known to be relevant reaction channels. For $^{10}$C projectile, inelastic scattering and breakup reaction could be of importance in the description of the elastic scattering on $^{208}$Pb target, even though the binding energies (for instance, 3.821~MeV for the $^{8}$Be+$p$+$p$) are relatively higher as compared to those for exotic nuclei, which are typically less than 1~MeV.

In the next sections the coupled-channels (CC) and the continuum-discretized coupled-channels (CDCC) calculations are explored to understand the effects of inelastic scattering and cluster structure in the elastic scattering of $^{10}$C.

\subsection{The Coupled-Channels calculations}

The coupling to the $2_1^+$ state in $^{10}$C has been show to be quite important to describe the elastic scattering of the  $^{10}$C+$^{58}$Ni system at $E_{\textnormal{lab.}} = 35.3$~MeV~\cite{GCS19}. Here, we assess this effect of couplings to the inelastic channel. The coupled-channels calculations were performed using the S\~ao Paulo potential (SPP) \cite{CCG02} for the real part and a short-range Wood-Saxon shape for the imaginary part of the optical potential. This short range imaginary potential takes into account the absorption flux due to the fusion process, since this channel is not explicitly included in the coupling scheme. The parameters for this potential are $W=50$ MeV, $r_W=1.06$ fm and $a_W=0.2$ fm, where $W$, $r_W$ and $a_W$ are the depth, reduced radius and diffuseness, respectively. This methodology has been applied with success to describe the elastic scattering of many systems \cite{CGL14,DGC15}. In the coupling scheme, the $2_1^+$ in $^{10}$C was treated in the independent model adopting the transition probability B(E2) from Ref.~\cite{RNT01}. The coupled-equations were solved using the \textsc{fresco} code \cite{Tho88}.

\begin{figure}[ht!]
\center
\includegraphics[width=0.5\textwidth]{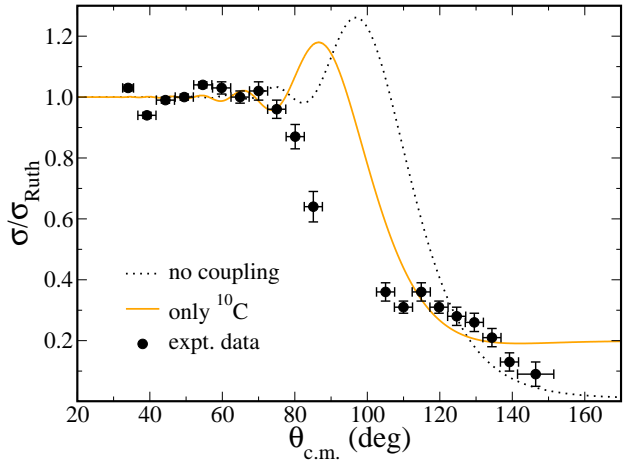}
\caption{Comparison between data and coupled-channels calculations for the elastic angular distribution for the $^{10}$C + $^{208}$Pb system at E$_{\rm lab}$ = 66 MeV. The no-coupling and the coupled-channel calculations are represented by dotted black and solid orange curves, respectively.}  
\label{cc}
\end{figure}

The comparison between data and coupled-channels calculations is shown in Fig.~\ref{cc}. The pronounced Fresnel peak present in the calculation is completely absent in the data. This is an indication that additional imaginary potential at the surface is necessary to reproduce the data.  The coupling to the strongly deformed $2_1^+$ state in the projectile (solid orange curve) is responsible for a shift and small reduction of the Fresnel peak and an increase of the elastic cross sections at backward angles compared to the no-coupling calculation (dotted black curve). This effect is also observed in the $^{10}$C+$^{58}$Ni system \cite{GCS19}. The inclusion of excited states of the target in the coupling scheme does not change significantly the results presented here (deviations less than 3\%).

The effect of transfer channels, such as neutron, proton and alpha, has been investigated for the $^{10}$C+$^{58}$Ni system \cite{GCS19} and little influence was observed. Here, we have considered the inclusion of one- and two-neutron pick-up reaction into the coupling scheme. The effect of such channels is small and only sizeable at large scattering angles and, therefore, not shown here.

\subsection{The three-body continuum-discretized coupled-channels calculations}

To investigate the effect of the continuum we performed continuum discretized coupled-channels calculations (CDCC). As mentioned before, the $^{10}$C nucleus may be described as $\alpha+\alpha+p+p$ cluster configuration. Such configuration would require a 5-body CDCC calculation (4 bodies in the projectile and the target nucleus), which is beyond the limits of contemporary reaction models. However, there is an indication that the breakup of $^{10}$C proceeds through sequential decays with three possible intermediate channels \cite{CAA08}: $^8$Be+$p$+$p$, $^{9}$B+$p$ and $^{6}$Be+$\alpha$ with binding energies of 3.821, 4.006 and 5.101 MeV, respectively. In this section we consider the effect of the two-body $^{9}$B+$p$, and $^{6}$Be+$\alpha$ cluster configurations in the CDCC calculations for the $^{10}$C + $^{208}$Pb elastic scattering. 

Although both $^9$B and $^6$Be cores are unbound ($^9$B $\rightarrow$ $\alpha$+$\alpha$+$p$ and $^6$Be $\rightarrow$ $\alpha$+$p$+$p$), we are implicitly assuming that these cores remain bound during the reaction and decay well away from the region of interaction with the target nucleus. This hypothesis is particularly an over-simplified picture in the case of $^6$Be, for which the ground state half-life is 92~keV ($\approx 7 \times 10^{-21}$~s) and comparable to the interaction time (roughly $10^{-21}$~s). However, it can be useful to assess the influence of two-body configurations in the scattering.

The projectile-target relative-motion wave functions were computed considering partial waves up to 1000$\hbar$ and radii up to $\mathit{R_{coupl}}$= 500 fm. The energy bins with relative angular momentum of the fragments up to $4\hbar$ were generated by the superposition of scattering states, and potential multipoles $\lambda$ up to the 5-pole term were considered. The energy bins were equally spaced in energy up to the maximum value of 10.0 MeV and width of 3.0 MeV, for the $^9$B+\textit{p} cluster configuration, and maximum value of 8.0 MeV and width of 3.0 MeV for the $^6$Be+$\alpha$ cluster configuration. These values were varied until a good convergence was achieved. 

To reproduce the structure of $^{10}$C in the two-body model, the $^9\text{B}+p$ and $^{6}\text{Be}+\alpha$ potentials are required. In both cases we have adjusted a Woods-Saxon potential to reproduce the energy and matter radius as obtained in the  three-body model for $^{10}$C (to be seen in the next subsection). In this way we can compare the results from the 3b- and 4b-CDCC. The shape of the adjusted potential is a Woods-Saxon form for central part with the radius fixed to 2.5 fm. For the $^{6}\text{Be}+\alpha$, the strength and diffuseness of the central part was 85 MeV and 0.8 fm. For the $^9\text{B}+p$ we also included a derivative Woods-Saxon potential for the spin-orbit. The strengths were 15 and 48 MeV for the spin-orbit and the central part, respectively. The diffuseness was 1.2 fm in both parts. The calculated ground-state matter radius are $r_{\rm m} = 2.73$~fm and $r_{\rm m} = 2.61$~fm for the $^9$B+\textit{p} and $^6$Be+$\alpha$ cluster configurations, respectively, for which it was assumed a radius of 0.87~fm for $p$ and 1.47~fm for $\alpha$. The SPP was used as the complex optical potential between the fragments and the $^{208}$Pb target. The real and imaginary parts were multiplied by a strength factor of 1.00 and 0.78, respectively.

\begin{figure}[ht!]
\center
\includegraphics[width=0.95\textwidth]{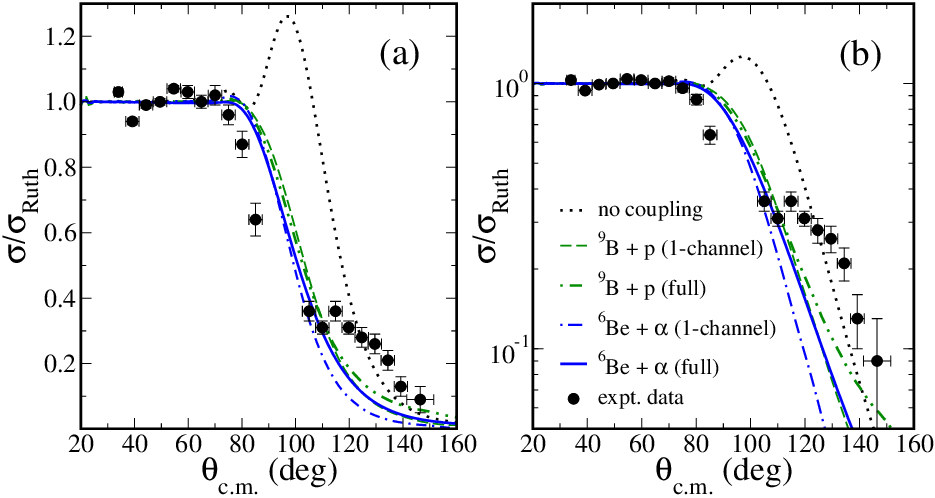}
\caption{Comparison between data and 3b-CDCC calculations considering the $^9$B + \textit{p} (green curves) and $^6$Be + $\alpha$ (blue curves) cluster configurations in (a) linear scale and (b) log scale. The no-coupling calculation, presented in the previous subsection, is also included (dotted black curve). See text for details.}
\label{cdcc_fig}
\end{figure}

The results of the 3-body calculations are presented in Fig.~\ref{cdcc_fig}. The dotted black curve corresponds to the calculation with no-couplings, using a projectile-target double folding potential, as discussed in the previous subsection. The dashed green and dot-dashed blue curves correspond to a folding cluster model for the $^{10}$C as one-channel $^9$B+$p$ and $^6$Be+$\alpha$, respectively, and considering only the ground state. In this model, couplings to the continuum states are switched off and the optical potential is defined as the expectation values of the sum of the fragments-target interactions over the projectile wave function. This optical potential is complex and the imaginary part is responsible for the suppression of the Fresnel peak. More details about this one-channel calculation for $^{10}$C can be found in Ref.~\cite{GCS19}. The inclusion of the $2_1^+$ state in $^{10}$C and continuum couplings are represented by solid blue and dot-dot-dashed green curves in Fig.~\ref{cdcc_fig}. The difference between 1-channel (no coupling to continuum) and full couplings is very small for both 3-body configurations. The major contribution of the full coupling, with respect to the one-channel calculation, comes from coupling to inelastic channel.

It is worth to highlight that, in this work, excited states are treated in a different fashion in the CC and CDCC formalisms. In the first, the deformation of the nuclei, which is the main ingredient to build the coupling potentials, is obtained within a collective model. In the CDCC formalism, excited states are built as single-particle states of a valence particle binding to its core.  We also checked the effect of different single-particle configuration in the $^9$B+$p$ cluster. In this case, the excited state can be build as the valence particle in the $1p_{1/2}$ or a $1p_{3/2}$. The results for elastic and inelastic cross sections are indistinguishable between these single-particle configurations. The 2-body description for the $2_1^+$ state results in a less deformed state and, consequently, its effect on the elastic angular distribution is attenuated compared to the CC calculation (see Fig.~\ref{cc}). In these cases, the deformation of the cores could be an important feature that shall be included in this type of calculations.

In Ref.~\cite{CAA08}, it has been suggested that $^9$B+$p$ is the dominant configuration even though the ground and excited states in $^{10}$C shall exhibit a mixture of cluster configurations. A full 3b-CDCC calculation, where both cluster configurations are simultaneously considered in the coupling scheme, is beyond the current possibilities. Keeping in mind that we are dealing with pure cluster configurations, the 3b-CDCC calculation considering the $^9$B+$p$ configuration gives a slightly better description compared to the $^6$Be+$\alpha$ channel, but the overall agreement between theoretical curves and the data is limited.

\subsection{\label{4b-cdcc} The four-body continuum-discretized coupled-channels calculations}

Going a step further in the CDCC calculations, we have performed four-body CDCC calculation  \cite{RAG08,RAG09}, in which the $^{10}$C is described within a $^8$B+$p$+$p$ configuration. The states of $^{10}$C are computed using a pseudostate approach with the analytical Transformed Harmonic Oscillator (THO) method presented in Ref.~\cite{JCasal13} for three-body projectiles. This formalism has been applied to reactions induced by the Borromean nucleus $^9$Be on different targets \cite{CRA15, ACR18}, showing the validity of the method to describe reactions induced by three-body projectiles with more than one charged particle.

To reproduce the structure of $^{10}$C in the three-body model, $^8\text{Be}+p+p$, the $^8\text{Be}+p$ and $p+p$ potentials are required. For the last, we have taken the Gogny-Pires-Tourreil (GPT) p-p potential \cite{gpt} whereas for the $^8\text{Be}+p$ subsystem we have adjusted a potential to reproduce the low-energy known states for $^9$B \cite{TILL04}, shown in table \ref{tab:levels}, for $p$ and $d$ waves. For $s$-waves we use the same as for $d$-waves. The shape of the adjusted potential is a Woods-Saxon form for central part and the derivative Woods-Saxon for the spin-orbit part. The values of the radius and the diffuseness were fixed to 2.5 fm and 0.65 fm. The spin-orbit strength was 22 MeV and the strength for the central part was 68.4 MeV for $s$- and $d$-waves and 38.5 MeV for $p$-waves. Apart from the two-body potentials, we include an effective three-body force to reproduce the energies of the bound states of $^{10}$C. The three-body force used has a Gaussian form depending on the hyper-radius $\rho$ (the three-body problem is treated in hyperspherical coordinates, see for example \cite{RAG08} for details): 
\begin{equation}
V_3(\rho)=v_3\exp{\left[\frac{\rho}{R_3}\right]}.    
\end{equation}
The radius $R_3$ was fixed to 5 fm, meanwhile the strength $v_3$ is adjusted to obtain the correct energies for bound states. This is fixed to -2.5 MeV for $j^{\pi}=0^+$ and 0.33 MeV for $j^{\pi}=2^+$. For $j^{\pi}=1^-$, we use the same strength as for $j^{\pi}=0^+$. To block the Pauli forbidden states we use the adiabatic projection method \cite{THOM00}.

\begin{table*} [t]
\caption{$^9$B energy states with respect to the $^8$Be$+p$ threshold.} 
\centering
\begin{tabular}{ c c  }
\hline
\hline
 $j^{\pi}$ & $E_x$ (MeV)  \\ 
\hline
  $3/2^-$ & 0.1851  \\
 $5/2^+$ & 2.5649    \\
 $1/2^-$&  2.6029   \\
\hline
\hline
\end{tabular}
\label{tab:levels}
\end{table*}

The parameters of the analytical THO basis chosen are fixed to $m=4$, $b=0.7$ and $\gamma=1.2$ (see Ref.~\cite{JCasal13}). The maximum hypermomentum is set to $K_{\rm max} = 8$, which has been checked to provide converged results for reaction calculations at the range of energies considered. The convergence is also reached using a THO basis with $i_{\rm max} = 8$ hyper-radial excitations. The calculated ground-state energy is $\varepsilon_B=-3.798$ MeV (fairly close to the binding energy 3.821~MeV) and rms matter radius $r_{\rm mat} =2.651$ fm, assuming a radius of 2.5 fm for $^8$Be.

Then, the $^{10}$C$-^{208}$Pb four-body wave functions are expanded in the internal states of the three-body projectile, leading to a coupled-equations system that has to be solved. For that, a multipole ($Q$) expansion is performed for each projectile fragment-target interacting potential. The procedure is explained in detail in Ref. \cite{RAG08}. The interaction between each projectile fragment and the target is represented by a complex optical potential, with both Coulomb and nuclear contributions. For consistency with the previous subsections, the complex optical potentials between the fragments and the $^{208}$Pb are the SPP with the strength factors indicated in the previous sections. In this way we can compare the results between the 3b-CDCC and 4b-CDCC on the same basis. Our model space includes $j^{\pi}=0^+,1^-,2^+$ states up to 8 MeV above the breakup threshold, which ensures convergence of the elastic angular distributions for this reaction. The coupled equations are solved up to 100 partial waves, including continuum couplings to all multipole orders, i.e., up to $Q =4$.

\begin{figure}[ht!]
\center
\includegraphics[width=0.5\textwidth]{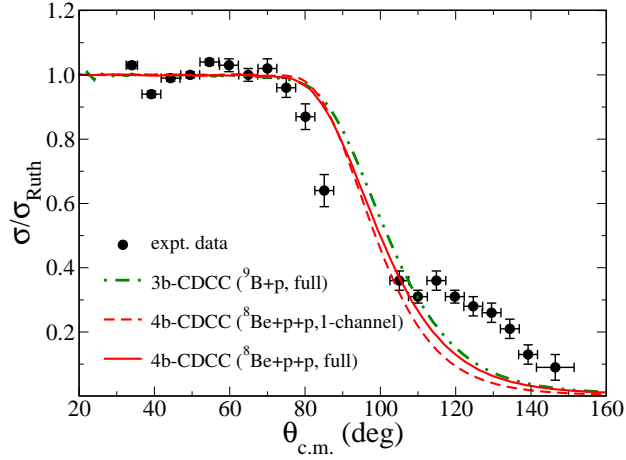}
\caption{Comparison between data and 4b-CDCC calculation considering the $^{10}$C nucleus has a $^8$Be+\textit{p}+\textit{p} cluster. The one-channel and full couplings results are represented by dashed and solid red curves, respectively. For reference, the no-coupling (dotted black curve) and the 3b-CDCC assuming the $^9$B+$p$ configuration (dot-dashed green curve) are also included.}
\label{fig:4bcdcc}
\end{figure}

The 4b-CDCC result for the elastic cross section is shown in Fig.~\ref{fig:4bcdcc}. The one-channel calculation is represented by the dashed red curve and the inclusion of couplings to the $2_1^+$ state and to the  continuum (solid red curve) just provides a small increase in the cross sections at backward angles, that is mainly due to the inelastic coupling. For the sake of comparison, we also included the full 3b-CDCC calculation within the $^9$B+$p$ configuration. The 3b- and 4b-CDCC results are quite similar and although both of them underpredicts the cross sections at the backward angles. The agreement with the data in the region of Fresnel peak is reasonable. We have to consider that a more realistic description of this nucleus would require a five-body model ($\alpha+\alpha+p+p$). Also, we stress that  double-folding SPP, that is adopted for the complex potentials, can be regarded as energy independent at energies around the coulomb barrier. The relevant feature for the calculations is its double-folding characteristic and there were no free adjustable parameters in these calculations. We have considered other optical potentials in the CDCC calculations and results are very similar.

\begin{figure}[ht!]
\center
\includegraphics[width=0.50\textwidth]{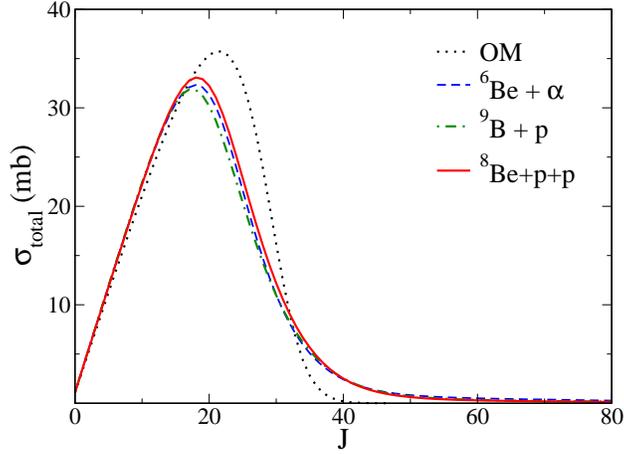}
\caption{Reaction cross sections as a function of the total angular momentum ($J$) as obtained from optical model fitting (OM), 3b-CDCC ($^6$Be+$\alpha$ and $^9$B+$p$) and 4b-CDCC ($^8$Be+\textit{p}+\textit{p}) calculations. }
\label{fig:sigmaJ}
\end{figure}

The comparison of reaction cross sections as a function of the total angular momentum of the scattering ($J$) between the OM, the 3b- and 4b-CDCC calculations is shown in Fig.~\ref{fig:sigmaJ}. The shape of the reaction cross sections (Fig.~\ref{fig:sigmaJ}) as obtained from the OM fitting (dotted black solid), the $^9$B+$p$ (dashed blue), the $^6$Be+$\alpha$ (dash-dotted green) and the $^8$Be+$p$+$p$ cluster (solid red) are all similar. We note that the maximum partial reaction cross section in the OM is at slightly higher $J$ ($\approx 21\hbar$) than in the CDCC calculations ($\approx 18\hbar$). The integrated cross sections are listed in Table~\ref{tab:CDCC_xsections} and are close to the value estimated in the optical model fitting ($\sigma_R$ = 753~mb). However, the behaviour of the reaction cross sections as obtained in the OM analysis is different than the ones obtained with the CDCC calculation.

\begin{table*} [t]
\caption{Total reaction cross sections ($\sigma_{\textnormal{total}}$) obtained from the OM and CDCC calculations.} 
\centering
\begin{tabular}{ c c }
\hline
\hline
model & $\sigma_{\textnormal{total}}$ (mb) \\ 
\hline
  OM                              & 753   \\
  $^{9}\textnormal{B} + p$        & 699   \\
  $^{6}\textnormal{Be} + \alpha $ & 715   \\ 
  $^{8}\textnormal{Be} + p + p$   & 727   \\
\hline
\hline
\end{tabular}
\label{tab:CDCC_xsections}
\end{table*}

\section{\label{conc}Conclusions}

The angular distribution for the elastic scattering of $^{10}$C on $^{208}$Pb target has been measured for the first time at an energy close to the Coulomb barrier. The experimental angular distribution was analyzed in terms of optical model, coupled channels and CDCC calculations. The total reaction cross section derived from the optical model analysis was compared to those from other projectiles, with similar masses, on the $^{208}$Pb target. The reduced reaction cross section is a little higher than those for $^6$He and $^8$Li projectiles in the same nuclei target. The enhanced reduced reaction cross section, compared to the UFF curve, and the absence of a Fresnel peak in the data indicate a strong absorption at the surface. 

From the coupled channel calculation, the couplings to the $2^+_1$ state in the $^{10}$C, treated as a collective state, is an important ingredient to describe the data at large scattering angles. The cluster configurations for the $^{10}$C has been studied within the 3b-CDCC, assuming the $^9$B+$\mathit{p}$ and $^6$Be+$\alpha$ configurations, and within the 4b-CDCC calculation where the $^{10}$C is described by the $^8$Be+$p$+$p$ configuration. Even though the binding energies of these configurations are high (more than 3.8~MeV), the calculations provide a reasonable description of the data, including the prediction of the suppression of the Fresnel peak but underestimate the data at backward angles. 

It would be interesting to consider a four-body configuration for the $^{10}$C which naturally will embrace clusters explored in this work and could provide a better description of the $2_1^+$ state. However, such calculation are still beyond the present capability of the reaction theory.

\section*{Acknowledgment}

The authors would like to thank the following foundations for financial support: CNPq (Grants 439375/2018-5 and 304961/2017-5),  S\~ao Paulo Research Foundation (FAPESP) (Grants 2016/02863-4 and 2016/17612-7), FAPERJ, CAPES (Grant PNPD 88887.475459/2020-00), INCT-FNA (Instituto Nacional de Ci\^encia e Tecnologia - F\'{\i}sica Nuclear e Aplica\c c\~oes) and Department of Energy, Office of Science, under Award No. DE-FG02-93ER40773. Also we would like to thank the operators team at the Cyclotron Institute of Texas A\&M University for reliable operation of the Cyclotron during the data acquisition that generated this work. This work was partially supported by the Spanish Ministry of Economy and Competitiveness, the European Regional Development Fund (FEDER), under Project $\rm N^o$  FIS2017-88410-P, by the European Union's Horizon 2020 research and innovation program, under Grant Agreement $\rm N^o$ 654002 and by SID funds 2019 (Universit\`a degli Studi di Padova, Italy) under project No.~CASA\_SID19\_01.
\pagebreak

%

\end{document}